\documentstyle[rmp,aps]{revtex}
\begin{document}
\title{\centering H\"older regularity and chaotic attractors}

\author{\centering Jacopo Bellazzini} 
\date{5 April 2001}
\maketitle
\begin{center}
\emph{Dipartimento di Ingegneria Aerospaziale dell'Universita' di Pisa\\
Via Caruso, 56100 Pisa, Italy}\\

(Received 4 April 2001)\\
\end{center}

\begin{center}
We demonstrate how the H\"older regularity of a given signal
is a lower bound for the Grassberger-Procaccia correlation dimension of
strange attractors
\end{center}

\\
It is known from the celebrated work by Lorenz [1] that even 
low-dimensional
deterministic dynamical systems may exhibit chaotic behavior. In the
context of turbulence in fluid dynamics, Ruelle and Takens [2] have shown
that the usual attractors of the asymptotic flow in the phase-space (fixed
points, periodic and quasiperiodic motion) cannot explain the sensitive 
dependence of the solutions  on the initial conditions . Attractors which 
display
chaotic features have been called by Ruelle and Takens ``strange attactors''.
Very often a strange attractor is a fractal object with different topological 
and metric (Hausdorff) dimensions. Grassberger and Procaccia [3] 
introduced 
the correlation dimension $\nu$ of strange attractors as a new measure related
to the fractal dimension \emph{D} by the relation $\nu < D$. The 
Grassberger and Procaccia method [3] is a practical algorithm to extract 
dimensional information from experimental data. Given a signal
${\vec X_{i=1,N}}={\vec X(t+\imath \tau)}_{i=1,N}$, the correlation integral is 
defined as the standard correlation function of the time series on the 
attractor:
\begin{equation}
c(r)=\frac {1}{N^{2}} \sum_{i,j=1}^{N} \theta(r-| \vec X_{i}-\vec X_{j} 
| ) \quad  ( N large) .
\end{equation}
where $\tau$ is the time step and $\theta (t)$ is the Heaviside function.
Grassberger and Procaccia have shown that \emph{c(r)} follows a power law of
\emph{r} for small \emph{r}:
\begin{equation}
c(r)=r^{\nu}
\end{equation}
\begin{equation}
 \nu =\lim_{r \rightarrow 0} \frac {ln(c(r))}{ln(r)}
\end{equation} 
In an experimental situation, for very small value of \emph{r} the poor 
statistics 
cause a scattering of the dimension $\nu$, but as \emph{r} increases  an 
interval $[r_{0},r_{1}]$ exists where the slope is constant. This region
is called the scaling region. The value of the slope in the scaling region 
is the
correlation dimension of the signal. 
The correlation dimension of a given signal is hence a function of the 
time step variable $\tau$. Let us suppose that the signal ${\vec X_{i}}$ is 
the 
discretization of a continuous function $f:I \rightarrow \mathbf{R}^{k}$ where
$I \subset \mathbf{R}$ is an interval where $f(t+\imath \tau)=\vec X_{i}$ for
$i=1,N$. The aim of this paper is to demonstrate that, given a large 
interval 
\emph{I} of fixed width $|I|$, in the limit of $N \rightarrow \infty$ and hence
$\tau \rightarrow 0$, the regularity of the function \emph{f} in terms of 
H\"older exponents gives a lower bound for the correlation dimension.\\
We suppose that \emph{f} is continuous in terms of the H\"older exponent over
the interval \emph{I}:
\begin{equation}
f \in C^{\alpha}(I)= \big\{  f \in C(I) : \forall \ t,t' \in I \,  \exists \,
\alpha \in  (0,1) \, \textrm{ and} \quad \exists d>0 : |f(t)-f(t')| 
\leq d|t-t'|^{\alpha} \big\}
\end{equation}
then the following proposition holds:\\
\\
\textbf{Proposition} \quad \emph{Let I be a large interval and} 
$f \in C^{\alpha}(I)$
\emph{with} $\alpha \in (0,1)$ \emph{and} $\big\{\vec X _{i}\big\}_{i=1,N}=
f(t+\imath\tau)$
\emph{with} $\tau =\frac {|I|}{N}$, \emph{then in the limit o}f $N \rightarrow \infty$ \emph{we have} $\nu \geq \frac{1}{\alpha}$\\
\\
Proof. We consider without any loss of generality the one dimensional case
$k=1$. Let $A(r)$ be the set where the Heaviside function is equal to 1
\begin{equation}
A(r)=\big\{ (t,t') \in \,  I \times I : |f(t)-f(t')|<r \big\}
\end{equation}
and
\begin{equation}
B(r)=\big\{ (t,t') \in \,  I \times I : |t-t'|<(\frac{r}{d})^{\frac{1}{\alpha}} \big\}.
\end{equation}
In the discrete case the set is \\
\begin{equation}
\tilde{B}(r)=\big\{ (i,j) \in \, [1,N] \times [1,N] : |i-j|<(\frac{r}{d})^{\frac{1}{\alpha}}\frac{1}{\tau} 
\big\}.
\end{equation}
Due to relation (4), $B(r) \subseteq A(r) \quad \forall r$ and so $\mu (B(r)) 
\leq \mu (A(r))$, where $\mu$ denotes   the usual Lebesgue measure. The 
cardinality of the set $B(r)$ is given by the relation $Card(B(r))=[\sqrt{2}N
(\frac {r}{d})^{\frac {1}{\alpha}}\frac{1}{\tau}]$, where  the symbol $[n]$ denotes
the greatest integer smaller than $n$. Without any loss of generality 
we assume  $Card(B(r))=\sqrt{2}N (\frac {r}{d})^{\frac {1}{\alpha}}\frac{1}{\tau}$ 
. The correlation integral is then greater than the following
quantity:\\
\begin{equation}
c(r) \geq \frac {\sqrt{2}}{N\tau} (\frac {r}{d})^{\frac {1}{\alpha}}.
\end{equation}
We can now evaluate the correlation dimension of the attractor:
\begin{equation}
\nu \geq \lim_{r \rightarrow 0} \frac{ln (\beta r^{\frac {1}{\alpha}})}{ln(r)}
\quad \beta= \frac {\sqrt{2}}{|I|d^{\frac {1}{\alpha}}}.
\end{equation}
We cannot choose $r$ arbitrarily small because for $(\frac {r}{d})^{\frac {1}
{\alpha}} \frac {1}{\tau} < 1$ the Heaviside function is zero for every 
couple
$(i,j) \in [1,N] \times [1,N]$. We should therefore consider  the limit 
\begin{equation}
\nu \geq \lim_{r \rightarrow (\tau)^{\alpha}d} \frac{ln (\beta r^{\frac 
{1}{\alpha}})}{ln(r)}.
\end{equation}
We consider $d=1$ without any loss of generality. Relation (10) implies 
that
\begin{equation}
\nu \geq \frac {1}{\alpha}(1+\frac {ln(\frac {\sqrt{2}}{|I|})}{ln(\tau)}).
\end{equation}
Given an arbitrarily small $\epsilon$, for small enough $\tau$ and hence 
sufficiently 
large
$N$ we obtain 
\begin{equation}
\nu \geq \frac{1}{\alpha} - \epsilon.
\end{equation}
\\

The significance of the stated proposition is that the regularity of a given 
signal
provides information about the topology of the attractor. It shows that the 
more
a given signal is non regular, in the sense of H\"older exponents, the greater 
is the number of dimensions  required to describe the 
attracting set in the phase-space. We cannot have an attractor with low 
dimensions in the phase-space
generated by a signal with strong singularities (low $\alpha$).
On the other hand, knowledge of the Grassberger-Procaccia correlation 
dimension of the attractor gives a lower bound for the regularity of
the signal. An obvious example is the 2-D classical Brownian
motion which we know  fills the full phase-space available, i.e its 
attractor is the 2-D plane. From this we can infer that its H\"older 
exponent has to be 
greater than  or equal to one-half, as it is [4].
Even if the propositon we have just demonstrated is quite general and applies 
to all non-linear
dynamical systems, a natural application can be found in fluid dynamics, 
where we will take as the phase-space the velocity-space.
 In
the case of the Navier-Stokes equations with the incompressibility condition
and with the initial boundary conditions on $\Omega \subset \mathbf{R}^{d}$, 
the 
main question concerns the existence and uniqueness of the solution.
The answer depends on the dimension $d$. In the case $d=3$ the answer is 
unknown due to the possible presence of singularities of the velocity field 
[5,6].
From the Kolmogorov statistical theory for isotropic and homogeneous 
turbulence [7], we know that the velocity in the limit of $Re \rightarrow 
\infty$ is not smooth but H\"older continuous of exponent one-third [8].
 The 
direct consequence of this H\"older regularity, which follows from the stated
proposition, is that the fractal dimension in the three dimensional
 velocity-space of the asymptotic attractor of a fully developed turbulent 
flow must be greater than or equal to 3, as it is.\\
 The author thanks G.Buresti and P.Burattini  
 for the continuous help in the preparation of the paper.

\end{document}